    \documentstyle[nato]{crckapb}

\input{psfig.sty}

\begin{opening}
\title{Chessboard magnetoconductance of a quantum dot in the Kondo regime}

\author{ C. Tejedor }
\institute{Departamento de F\'{\i}sica Te\'orica de la Materia Condensada,
\\ Universidad Aut\'onoma de Madrid, \\ Cantoblanco, 28049, Madrid, Spain.}

\author{ L. Martin-Moreno }
\institute{ Departamento de F\'{\i}sica de la Materia Condensada, 
\\ Universidad de Zaragoza, \\ Zaragoza 50015, Spain.} 

\end{opening}

\runningtitle{THE CRCKAPB STYLE FILE}

\begin{document}


\par

\begin{abstract}
Transport through a quantum dot (QD) in the Kondo regime
shows alternating regions of high and low conductance when 
both an external magnetic field and the gate potential 
controlling the depth of the QD potential are varied. 
We present a theoretical analysis of this chessboard aspect of the 
magneto-conductance. An effective Kondo Hamiltonian is obtained by 
means of a restriction to the Hilbert space supported by just 
a few low energy states of $N$ and $N \pm 1$ electrons in the QD. 
We obtain antiferromagnetic exchange couplings depending on tunneling 
amplitudes and correlation effects. When either the magnetic field or the 
number of electrons in the QD is varied, Kondo temperature shows large 
oscillations due to the successive appearance of ground states having strong 
and weak correlation effects alternatively. 
\end{abstract}

\section{Introduction}
Many theoretical predictions\cite{Glazman,Ng,Kawabata,Hershfield,Meir,Levi,Matveev} 
on the Kondo physics in a quantum dot (QD) have been experimentally 
observed\cite{Goldhaber,Cronenwett,Stutt1,Simmel,Delft,Delft2}. 
However, the use of a standard Anderson Hamiltonian does not allow a complete 
understanding of some other experiments as those devoted to measure the phase-shift 
of electrons passing through the QD\cite{Ji1,Ji2} or those presenting a chessboard 
aspect of the QD conductance in the quantum Hall regime\cite{Stutt2,Stutt3,Sprinzak}. 
Correlation effects within the QD\cite{Tejedor,Silvestrov} are 
probably essential for the understanding of the experimental results.

This paper is devoted to the theoretical analysis of the experimental Kondo-like 
conductance of a QD in the quantum Hall regime. For a given number $N$ 
of electrons in the QD, the system presents alternating high and low conductance 
regions as a function of an external magnetic field $B$. When $N$ is varied in 
$\pm 1$, the high and low conductance valleys are interchanged. Therefore, the 
representation of the conductance (in a grey scale) as a function of both $B$ 
and a gate potential which allows to vary $N$, takes 
the aspect of a chessboard\cite{Stutt2,Stutt3,Sprinzak}.
An interesting piece of experimental information on the density of states is 
obtained from the differential conductance measured when a source-drain bias 
is applied. In such experiment\cite{Stutt2,Stutt3},
one observes a double peak with a separation which remains practically constant 
when the magnetic field is varied. This suggests that two (split) states 
are playing an important role in the conduction process. 

Our description for the Kondo physics in a QD in the presence of a high magnetic 
field is valid for any number of electrons (even or odd without restriction to 1 
or 2) and any value of the QD spin $S_z$ (not restricted to 0, 1 for even $N$ 
or 1/2 for odd $N$).
Instead of describing spin-flip scattering in terms of spin-ladder operators 
$S^{(\pm)}$, we find a set of spin-flip 
Hubbard operators describing a collective spin effect of all the $N$ electrons 
contained in the QD. Despite both $N$ and $S_z$ 
can be very large, the scattering of carriers only produces transitions 
between two many-body states of the QD with $S_z$ differing in $\pm 1$.
These many-body states are of two types: some of them are "practically" just one 
Slater determinant, i.e. not presenting correlation effects, while some others 
are a linear combination of many configurations, i.e. involving strong correlation 
effects. This is going to produce strong oscillations in the Kondo temperature. 
We find this is the physical reason of the appearance of the chessboard aspect of the 
conductance instead of the other obvious possibility of having alternating ferro and 
antiferro exchange couplings. In fact, we find that the spin-flip process can 
be described by a Kondo Hamiltonian with antiferromagnetic couplings which depend 
on both tunneling amplitudes and correlation effects. 

The steps of our description are the following:
\begin{enumerate}

\item Hamiltonian for N interacting electrons in a QD coupled to two leads.

\item Numerical calculation of the spectrum of N electrons in the isolated QD
which involves three energy scales: confinement, Zeeman and interaction.

\item Two alternating types of ground states (GS): strong correlation and weak 
correlation.

\item Projection onto the subspace of the (two) lowest eigenstates and 
coupling to the leads.

\item An effective Kondo Hamiltonian with antiferromagnetic exchange couplings 
depending on the QD properties (Spectral amplitudes) is obtained.

\item Strong oscillations of the Kondo temperature due to strong oscillations
of the spectral amplitude for strongly and weakly correlated states.

\item Experiment at finite temperature $\Rightarrow $ oscillations (Chessboard
aspect) of the conductance.

\end{enumerate}

\section{QD spectrum (Steps 1 to 3)}
We consider $N$ electrons in the presence of a magnetic field and confined in a 
QD coupled to leads. The Hamiltonian is 
\begin{eqnarray}
H=H_{QD}+H_{L}+H_{TUN} . 
\label{hamilt}
\end{eqnarray} 
The first term in Eq. (\ref{hamilt}) describes an isolated QD. For a parabolic 
QD in the presence of a magnetic field and including interaction between 
the electrons, the QD Hamiltonian is described by 
(hereafter we take $\hbar =1$)
\begin{eqnarray}
H_{QD} = 
\sum _{i} \varepsilon _{i,\sigma} d^\dagger_{i,\sigma} d_{i,\sigma}
+ g \mu_B BS_z + \frac{1}{2} \sum_{i_j,\sigma _j} V_{i_1i_2i_3i_4}
d^{\dagger}_{i_1,\sigma_1} d^{\dagger}_{i_2,\sigma_2}
d_{i_3,\sigma_3}d_{i_4,\sigma_4} .
\label{hqd}
\end{eqnarray}
The first term in (\ref{hqd}) involves the energy scale related with potential and 
magnetic confinement. $i=\{n,m\}$ is an index containing both the Landau level index
$n$ and the third component of the angular momentum $m$.
\begin{eqnarray}
\varepsilon _{i,\sigma}=\left[ \frac{1}{2}+\frac{1+\gamma}{2}n+\frac{1-\gamma}{2}m
\right] \Omega
\label{energ}
\end{eqnarray}
is the single particle eigenvalue depending on  
both the QD confinement $\omega _0$ and cyclotron $\omega _c$ frequencies
through $\Omega=\sqrt{\omega _c^2+4\omega _0^2}$ and $\gamma =\omega _c/\Omega$. 
$d^\dagger _{i,\sigma}$ creates an electron with Landau index $n$, angular
momentum $m$ and spin $\sigma $ in the QD.
The second term in Eq. (\ref{hqd}) introduces the Zeeman energy scale 
with $g$ being the Land\'e $g$-factor.
The interaction energy scale appears in the last term, electron-electron repulsion.  
The Coulomb interaction matrix 
elements $V_{i_1i_2i_3i_4}$ have a typical energy scale 
$e^2/\varepsilon l_B$, where $\varepsilon $ is the dielectric constant and 
$l_B=1/\sqrt{m\Omega}$ the magnetic length. 

The second term in Eq. (\ref{hamilt})
\begin{eqnarray}
H_{L}=\sum _{k,\sigma} \varepsilon _{k,\sigma} c^\dagger_{k,\sigma} c_{k,\sigma}
\end{eqnarray}
describes, in a single-particle approach, the leads having electrons 
with quantum numbers $k,\sigma$ 
occupying states up to the Fermi energy $\varepsilon _F$. 

Finally the tunneling part of the Hamiltonian (\ref{hamilt})
\begin{eqnarray} 
H_{TUN}=\sum_{k,i,\sigma}V_i \left( d_{i,\sigma} ^\dagger c_{k,\sigma}+
c ^\dagger _{k,\sigma} d_{i,\sigma} \right)
\label{htun}
\end{eqnarray}
is written without any dependence on $k$ of the tunneling amplitudes $V_{i}$ 
(taken as real positive) but retain the dependence on $i$ because 
is going to produce physical consequences.
\begin{figure}
\hspace{2.5cm} \psfig{figure=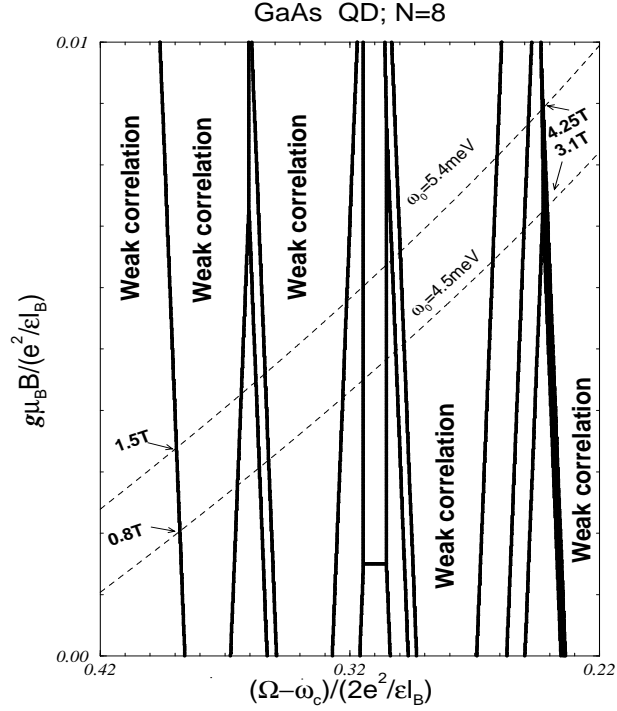,height=10.0cm,width=8.0cm}
\caption{Phase diagram of the possible GS of a GaAs QD with 8 electrons 
at $2>\nu>1$. The axes contain kinetic 
$(\Omega-\omega _c)/(2 e^2/\varepsilon l_B)$ and Zeeman 
$g\mu _B B/(e^2/\varepsilon l_B)$ contributions to the energy.
The regions labelled as {\it weak correlation} correspond to compact 
states from $|C_4^4\rangle $ ($\nu =2$) at the left
to $|C_8^0\rangle $ ($\nu =1$) at the right. 
The unlabelled regions correspond to GS showing strong correlation effects 
(skyrmion-like states, see text). 
Dashed lines depict, for two different values of the QD 
confinement frequency $\omega _0$, the evolution of the GS within the range 
of magnetic fields given at the edges of the lines.}
\label{fig1}
\end{figure}

$H_{QD}$ can be numerically diagonalized for a significantly broad 
range of $N$ and $B$, as discussed in many theoretical papers\cite{Palacios}. 
We have performed calculations including $n=0,1$ which means we are able 
to analyze filling factors $\nu$ below 4, i.e. the regime of experimental 
interest \cite{Stutt2,Stutt3}. Fig. \ref{fig1} shows results for the GS of $N=8$ for 
$B$ large enough to have electrons occupying states only with the lowest Landau 
index $n=0$, i.e. $\nu \leq 2$. 
Dashed lines in Fig. \ref{fig1} depict, for two different values of $\omega _0$, 
the evolution of the GS for a GaAs QD within the range of magnetic fields 
(perpendicular to the QD) given at the edges of the lines. 
The first observation is that there are two type 
of regions: 5 regions, labelled as weak correlation, correspond to GS which are 
practically just one single Slater determinant (or configuration). The other 
regions correspond to GS which are a linear combination of many configurations. 
In other words, there are regions with GS having weak correlation and regions with GS 
having strong correlations. This trend also occurs for calculations in other 
situations of $N$ and $B$. In the particular case of Fig. \ref{fig1}, GS's in the 
weak correlation regions are compact states of the form 
\begin{eqnarray}
|C^K_{N-K} \rangle =\prod _{m=0} ^{K-1} d^\dagger _ {n=0,m,\downarrow} 
\prod _{m=0} ^{N-K-1} d^\dagger _{n=0,m,\uparrow} |0 \rangle 
\end{eqnarray}
where $|0 \rangle$ is the vacuum state.
In going from left to right, one finds successively the states 
$|C_4^4\rangle $ ($\nu =2$), $|C_5^3\rangle $, $|C_6^2\rangle $, 
$|C_7^1\rangle $ and $|C_8^0\rangle $ ($\nu =1$).
In the case of Fig. \ref{fig1}, strongly correlated states have a very peculiar 
shape: they are skyrmion-like states of topological charge 
1 \cite{Tejedor,Oaknin1,Oaknin2}. While the occurrence of compact states is general 
in the calculations we have performed, in many of such calculations we have not been 
able to identify the particular nature of the highly correlated GS that always exist. 

\begin{figure}
\hspace{2.5cm} \psfig{figure=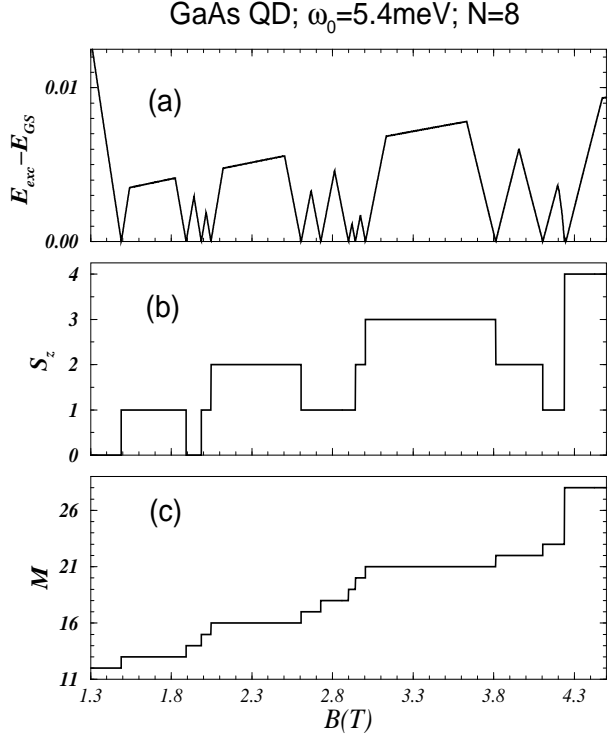,height=10.0cm,width=8.0cm}
\caption{Evolution with the magnetic field $B$ of different magnitudes for 
the GS of a GaAs QD with $N=8$ and $\omega_0=5.4$meV.
Energies are measured in units of $e^2/\varepsilon l_B$. The left 
side of the figure corresponds to $\nu=2$ and the right side to $\nu =1$.}
\label{fig2}
\end{figure}

Fig. \ref{fig2} shows the evolution of the GS properties with increasing $B$
for the system of Fig. \ref{fig1} for a typical value \cite{Ashoori} $\omega _0=
5.4$meV. The GS energy $E_{GS}$ has a kink any time a crossing 
of states occurs. The relevant information on these crossings is 
the energy splitting $\Delta E=E_{exc}-E_{GS}$ between the lowest excited state 
and the GS as shown in part (a) of the figure. The spin $S_z$ and the third 
component $M$ of the total angular momentum of the GS are also given in 
(b) and (c) respectively. The general trend of interest for us is that GS changes 
many times when $B$ is varied. For many values of the field, the energy 
difference between the two lowest states is very small. 
Moreover, in almost all the cases, these two states
have spins differing in 1 while they have different $M$. 

\section{Kondo Hamiltonian (steps 4 to 5)}
As we mentioned in the introduction, experiments \cite{Stutt2,Stutt3} show 
that the density of states presents a double peak very stable with $B$.
As a consequence of both this experimental fact and the above numerical results, 
we consider that the main physics
of the problem is captured by a two level system approach \cite{Cox}.
We project the problem of a $N$ electrons in a QD affected by a high magnetic 
field onto the subspace subtended by the two lowest energy states
$|N,\Uparrow \rangle$ and $|N,\Downarrow \rangle$ with $M_\Uparrow 
\neq M_\Downarrow $ and spins differing in 1, i.e.  $\langle N,\Uparrow 
|S_z|N,\Uparrow \rangle= \langle N,\Downarrow |S_z|N,\Downarrow  
\rangle +1$. $H_{TUN}$ mixes these states with states $|N\pm1 \rangle $ 
in which $N\pm 1$ electrons are in the QD. In our two level system description, 
the tunneling Hamiltonian is projected onto the subspaces subtended by the two 
states for $N$ electrons and the connecting $|N\pm1 \rangle$ electron states. 
In this process, we consider the connecting $|N\pm1 \rangle$ states as 
non-degenerate which is the most common 
case. If either $|N+1\rangle $ or $|N-1\rangle $ is also 
degenerate, the algebra is more complicated but the physics is the same. 
Using the notation $\Sigma \equiv \{ \Uparrow , 
\Downarrow \}$, we introduce the tunneling spectral amplitudes  
\begin{eqnarray}
& & \Delta_{-,\Sigma}= \sum_{i,\sigma} V_i \langle N-1|d_{i,\sigma}|
N,\Sigma  \rangle 
\nonumber \\ 
& & \Delta_{+,\Sigma}= \sum_{i,\sigma} V_i \langle N+1
|d^\dagger_{i,\sigma}| N,\Sigma \rangle \, 
\label{delta}
\end{eqnarray}
and the spin-flip Hubbard operators
\begin{eqnarray}
X_{\Sigma, \Sigma'}=|N,\Sigma\rangle \langle N,\Sigma' |.
\label{hubbards}
\end{eqnarray}
The projected tunneling Hamiltonian $\overline{H}_{TUN}$ describes the interaction 
between the QD with $N$ electrons and the leads. An effective coupling is obtained 
by a standard scattering description \cite{Hewson} including all the possible 
intermediate states $|I\rangle$ up to second order: 
\begin{eqnarray}
& H_{eff} & = 
\sum_{I} \frac{\overline{H}_{TUN}|I\rangle \langle I|
\overline{H}_{TUN}}{E_{GS}^N-E_I} 
\nonumber \\
& & = \sum_{k,k',\Sigma} \! \left[   
\frac{\Delta_{-,\Sigma} ^2 \delta_{k,k'}X_{\Sigma,\Sigma}}
{E^N- \! E^{N-1}- \! \varepsilon_F} 
+ \frac{\Delta_{+,\Sigma}^2c^\dagger_{k',\sigma}c_{k,\sigma}}
{E^N- \! E^{N+1}+ \! \varepsilon_F} \!
\right. \nonumber \\ 
& & \left. + \! \! \sum _{\Sigma'} \! \left(\frac{\Delta_{+,\Sigma}\Delta_{+,\Sigma'}}
{E^{N+1} \! - \! \varepsilon_F-\! E^N}+
\frac{\Delta_{-,\Sigma}\Delta_{-,\Sigma'}}{E^{N-1}+
\! \varepsilon_F-\! E^N} \! \right) \!  
X_{\Sigma,\Sigma'} c^\dagger _{k',\sigma'}c_{k,\sigma} 
\right] ,
\label{heff}
\end{eqnarray}
where we have neglected the energy difference between the two states of $N$ electrons.
Due to spin conservation, $\Sigma$ has the same direction than 
$\sigma$ and $\Sigma'$ the same than $\sigma'$.
The first term is simply a constant. The second term (involving $ 
c^\dagger_{k',\sigma}c_{k,\sigma}$) represents a potential scattering 
which does not involve any spin flip. These two terms are identical 
to the ones appearing when building up an $sd$ Hamiltonian from the 
Anderson Hamiltonian in the case of $N=1$ \cite{Hewson}. 
As it is usually done in that case, one can forget 
about these two terms which do not contain anything important for the 
physics we want to address. 

The interesting physics is included in the third term of (\ref{heff}) 
which is a Kondo Hamiltonian
\begin{eqnarray}
H_K= \sum_{k,k'} \left[ 
J\left( X_{\Uparrow,\Downarrow} c^\dagger _{k'\downarrow}c_{k,\uparrow} 
+X_{\Downarrow,\Uparrow} c^\dagger _{k', \uparrow}c_{k,\downarrow} \right) 
+J_{\Uparrow} X_{\Uparrow,\Uparrow}c^\dagger _{k', \uparrow}c_{k,\uparrow}
+J_{\Downarrow} X_{\Downarrow,\Downarrow} c^\dagger _{k', \downarrow}c_{k,\downarrow} 
\right] ,
\label{kondo}
\end{eqnarray}
with exchange couplings
\begin{eqnarray}
J = \frac{\Delta_{+,\Uparrow}\Delta_{+,\Downarrow}}
{E^{N+1}- \varepsilon_F-E^N} +
\frac{\Delta_{-,\Uparrow}\Delta_{-,\Downarrow}}{E^{N-1}+
\varepsilon_F-E^N} 
\label{exchange}
\end{eqnarray} 
\begin{eqnarray}
J_{\Sigma} = \frac{\Delta_{+,\Sigma}^2}{E^{N+1}-
\varepsilon_F-E^N} + \frac{\Delta_{-,\Sigma}^2}{E^{N-1}+ 
\varepsilon_F-E^N}.
\label{exchangesigma}
\end{eqnarray} 
$H_K$ is a spin-flip scattering Hamiltonian in which the the two possible states 
of the scatterer flip their spins by means of $X_{\Uparrow,\Downarrow}$ and 
$X_{\Downarrow,\Uparrow}$ and, at the same time, 
they change their total angular momentum. Both the difference between $M_\Uparrow $ 
and $M_\Downarrow$, and the correlation effects included in the tunneling 
spectral amplitudes, are the reason why one must use spin-flip Hubbard operators 
instead of the usual spin-ladder operators $S^{(\pm)}$. 

A crucial question is the sign of the exchange couplings (\ref{exchange}) and 
(\ref{exchangesigma}). With our definitions (\ref{delta}), all the tunneling 
spectral amplitudes are positive. The sign of 
$\Delta_{+,\Uparrow}$ is the same than the one of $\Delta_{+,\Downarrow}$, and 
the same happens for $\Delta_{-,\Uparrow}$ with respect to $\Delta_{-,\Downarrow}$.
Therefore, the signs of the exchange couplings are determined by the denominators.
As, in the considered situation, the lowest energy corresponds to having 
$N$ electrons in the QD, all the intermediate 
states $|I\rangle$, with $N \pm 1$ electrons in the QD and $\mp 1$ 
electron at the Fermi level of the leads, have higher energy. 
Therefore, we have a very 
important result: {\it $H_K$ has positive effective exchange couplings}. 
So, a Kondo Hamiltonian (\ref{kondo}) with {\it antiferromagnetic couplings} 
(\ref{exchange}) and (\ref{exchangesigma}) has been obtained. The conclusion is: 
{\it for any values of both $N$ and $S_z$, the QD in the the presence of high $B$,
presents Kondo physics}. 

\section{Exchange couplings and Kondo temperature (step 6)}
The Hamiltonian (\ref{kondo}) is equivalent to the standard 
Kondo Hamiltonian for $N=1$. Any other measurable property deduced from 
Hamiltonian (\ref{kondo}) as the temperature dependence of the conductance must 
present characteristics similar to the one found in the experiments involving just 
one electron \cite{Goldhaber,Cronenwett,Stutt1,Simmel}. 

The important issue to be 
discussed in the general case of any $N$ is the characteristic energy scale, $T_K$, 
which is determined by the antiferromagnetic couplings. 
$J$ and $J_{\Sigma}$ depend on both the energy difference 
$E^{N}-E^{N \pm 1}\mp \varepsilon_F$, 
and tunneling spectral amplitudes $\Delta_{\pm ,\Sigma}$. The 
former is practically independent on both $N$ and $\Sigma$. 
Therefore, it does not imply any significant difference with respect to the 
well known case of $N=1$. However interesting physical differences appear, 
through $\Delta _{\pm ,\Sigma}$, depending on the weak or strong correlation 
nature of the states  $|N,\Sigma \rangle$. 
Let us analyze the different situations as far as correlations are concerned:

{\it i)} {\it Weak correlation}

In order to simplify the discussion, let us consider the simple case of Fig. 
\ref{fig1} where the weak correlation states are compact states 
\begin{eqnarray}
|N, \Uparrow \rangle=|C^K_{N-K} \rangle  \, \, \, ; \, \, 
|N, \Downarrow \rangle=|C^{K+1}_{N-K-1}\rangle .
\end{eqnarray}
This is the simplest case in which 
\begin{eqnarray}
X_{\Uparrow,\Downarrow}=(-1)^K d^\dagger _ {n=0,N-K-1, \uparrow}d_ {n=0,K, \downarrow}
\end{eqnarray}
and similarly for the other Hubbard operators.
We have also assumed, $|N-1\rangle= 
|C^{K}_{N-K-1} \rangle$ and $|N+1\rangle=|C^{K+1}_{N-K} \rangle$.
In other case the quantum numbers of the 
$d^\dagger$ operators must be changed accordingly. Moreover, the signs are taken 
according with definitions (\ref{hubbards}). Due to the lack of correlation 
effects in the states, the tunneling spectral amplitudes are 
\begin{eqnarray}
& & \Delta_{-,\Uparrow}= \Delta_{+,\Downarrow}=V_{N-K-1} \, ; \, 
\Delta_{-,\Downarrow}=\Delta_{+,\Uparrow}=V_{K}.
\end{eqnarray}
There is a difference between the antiferromagnetic 
couplings $J$, $J_\Uparrow$ and $J_\Downarrow$ due to the fact that $N-K-1 
> K$. It is easier to tunnel from the leads to spin up states 
($m=N-K-1$) within
the QD because they are in the outer region of the QD while the first 
available spin down state ($m=K$) is in the inner region of the QD. 
This tunneling amplitude effect is not described by previous models
of Kondo at finite B for $N=1,2$ \cite{Pustilnik,Eto,Giuliano}, but has long been 
more broadly recognized both experimentally and theoretically for QD in the 
quantum Hall regime.

In this weak correlation regime the QD does not present any correlation effects, but 
the antiferromagnetic couplings are still a function of total energies and 
tunneling amplitudes. Since the Kondo temperature depends exponentially on 
$J$\cite{Hewson}, it is a very sensitive magnitude with respect to many 
parameters as $\omega _0$, $B$, $V_i$ etc. However, this is not 
different from the case of just one electron in the QD in which experiments 
\cite{Goldhaber,Cronenwett,Stutt1,Simmel} show $T_K$ to be in the range of 1K.
In any case, one can predict variations in $T_K$ when 
the regime changes as discussed below. 

{\it ii)} {\it Strong correlation}

In this case the discussion is much more complicated because for magnetic fields 
implying $\nu \geq 2$ we have not been able to find an analytical expression for 
the states. Therefore, we are going to make the discussion restricting us to the 
particular case of Fig. \ref{fig1} in which the strongly correlated states 
skyrmion-like states of topological charge one \cite{Oaknin1,Oaknin2}. In this case,
The Hubbard operators $X_{\Sigma, \Sigma '}$ have 
complicated, but analytical, expressions \cite{Tejedor}.

In some of the experiments showing chessboard conductance \cite{Stutt3} $N$ is 
of the order of 50. For such a large number of electrons,
correlation effects provoke that $\Delta _{-,\Sigma}$ tends to zero. For instance, 
$\Delta _{-,\Sigma} \propto [N ln N]^{-1/2}$ for the skyrmion with the 
smallest size. This implies a quenching of the antiferromagnetic couplings due to the 
orthogonalization catastrophe. As a consequence, Kondo effect should not 
be observed when the number of electrons within the QD is large because 
the Kondo temperature is extremely small in this case. 

In some other experiments showing chessboard conductance \cite{Stutt2} $N$ is lower 
than 10. In this case,  
correlation effects do not destroy Kondo effect but reduce $\Delta_{-,\Sigma}$ 
up to a factor of two \cite{Palacios}. The reduction of $J$ and $J_{\Sigma}$ 
implies that Kondo temperatures are significantly smaller 
than for the case {\it i)} of compact states. In practice, one can move from 
the strong correlation regime {\it ii)} (lower part of Fig \ref{fig1}) to the 
weak correlation regime  
{\it i)} (upper part of Fig \ref{fig1}) by increasing an inplane 
component of the magnetic field. As a consequence of the above analysis, 
one should detect a clearly measurable increase of $T_K$ during this process.

{\it iii)} There is a rather unusual case, for instance in the last step to 
the right in Fig. \ref{fig2}, in which the two lowest energy states with $N$ 
electrons have spins differing in more than 1. Therefore, 
they can not be obtained from the same $|N-1 \rangle $ by creating  
one electron with spin either up or down.
In other words, one of the two states has a tunneling spectral amplitude equal 
to zero so that, for tunneling effects there is only a non degenerate GS 
and Kondo effect does not occur. 

\section{Chessboard behavior of the conductance (step 7)}
A very characteristic feature of some experiments \cite{Stutt2,Stutt3,Sprinzak} is 
an alternating high-low conductance sequence as a function of $B$ for a 
given temperature and number of electrons. When $N$ is varied in $\pm 1$, 
the high and low conductance regions are interchanged. Since a
common way of representing the experimental results is to use a color-intensity 
scale for the conductance as a function of both $B$ and a gate potential which 
varies $N$, the data present a chessboard aspect\cite{Stutt2,Stutt3,Sprinzak}.
This occurs in a broad range of filling factors and number of electrons.   

\begin{figure}
\hspace{2.5cm} \psfig{figure=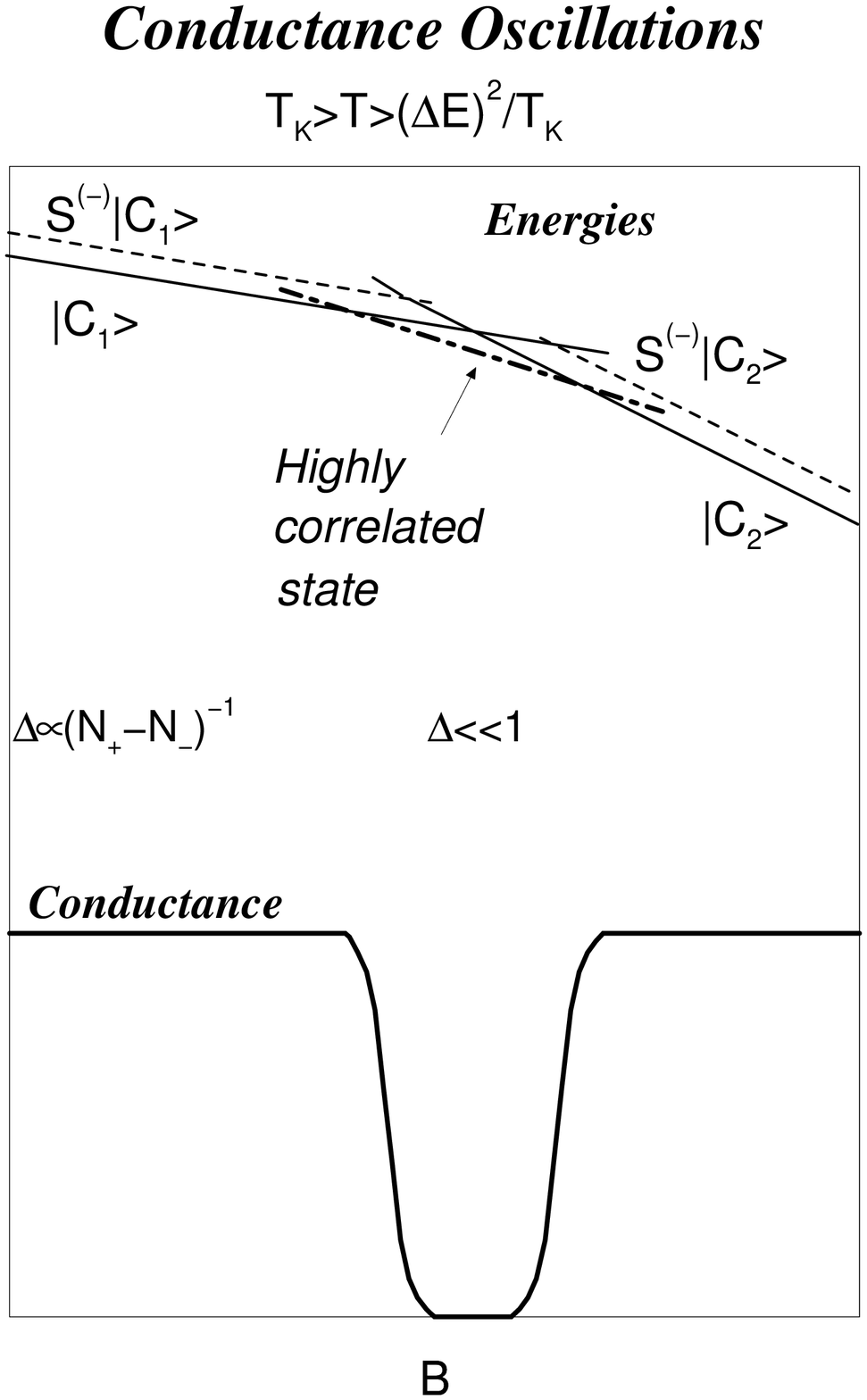,height=10.0cm,width=8.0cm}
\caption{Schematic evolution with the magnetic field of the lowest
energy states of $N$ electrons within a QD (upper part). Continuous and 
dashed lines correspond to weak correlation states, while dashed-dotted 
lines correspond to strong correlation states. Kondo-like 
conductance (lower part) through such a QD in the
range $T_K > T > \Delta E^2/T_K$ (see text).} 
\label{fig3}
\end{figure}

In the two level system approach\cite{Cox} in which the two lowest states
are separated by an energy splitting $\Delta E$, Kondo-like  
behavior appears only when the experimental temperature is between
$(\Delta E)^2/T_K$ and $T_K$ (provided $\Delta E < T_K$) 
in order to have significant occupation of the two states. 
The quantitative application of the model would require the computation of 
both the energy splitting $\Delta E$ and Kondo temperature $T_K$ which
are very sensitive to many experimental parameters. Instead, a 
general understanding of the chessboard is obtained from the following
qualitative explanation: 
as mention before, the main ingredient of our description is the alternation, 
when varying $B$, of weak and strong correlated GS of $N$ within the QD. 
The second crucial point is that for weak correlation states, spectral amplitudes 
are much higher than for strong correlated states. These two features are 
depicted in the upper part of Fig. \ref{fig3} where we have labelled the weak 
correlated states as compact states $|C_{1,2}\rangle$ while we have not given any 
name to the strongly correlated state. The important consequence of the 
alternation of high and low spectral functions is the alternation of high and low 
exchange couplings and consequently the same alternation of Kondo temperatures $T_K$
which depends exponentially on the exchange coupling \cite{Hewson}.
Since the experiment is performed by fixing a temperature $T$, the strong oscillations 
of $T_K$ imply that $T$ alternates being lower (for weak correlation regions) and 
higher (for strong correlation regions) than $T_K$. In other words, 
$T<T_K$ for weak correlation regions, which means a high Kondo-like conductance. 
On the contrary, $T>T_K$ for strong correlation regions, which implies a quenching 
of the conductance. This is depicted in the lower part of Fig. \ref{fig3}. 
This explains the alternating behavior experimentally observed for fix gate 
voltage (i. e. number of electrons in the QD) and varying magnetic field.

\begin{figure}
\hspace{2.5cm} \psfig{figure=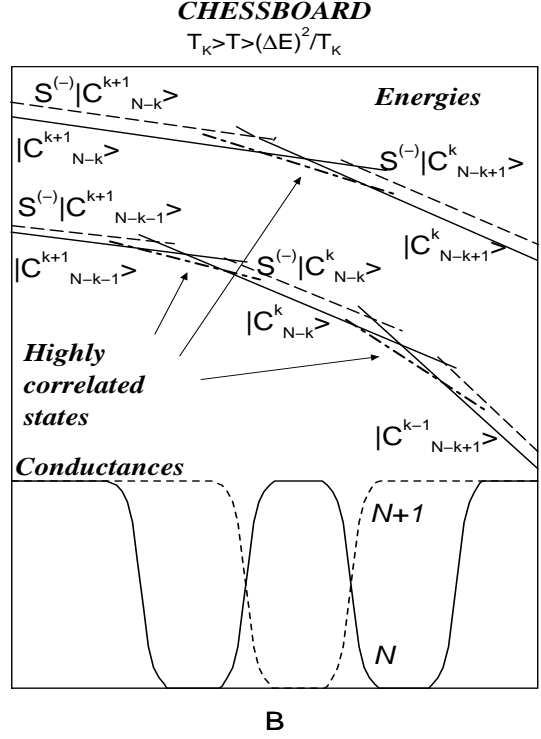,height=10.0cm,width=8.0cm}
\caption{Schematic evolution with the magnetic field of the lowest
energy states of $N$ and $N+1$ electrons within a QD (upper part). Continuous 
and dashed lines correspond to weak correlation states, while dashed-dotted
lines correspond to strong correlation states. 
Kondo-like conductance (lower part) through such a QD in the
range $T_K > T > \Delta E^2/T_K$. The oscillations of the conductance are shifted 
between $N$ and $N+1$ cases due to the dephasing between the positions of the 
crossing of GS in the two cases.}
\label{fig4}
\end{figure}

The chessboard aspect also implies alternating low-high conductance  
regions for fix $B$ and varying gate voltage. This is due
to the fact that crossings for $N\pm 1$ electrons occur for magnetic fields 
roughly midway from crossings for $N$ electrons\cite{Palacios} as depicted 
qualitatively in Fig. \ref{fig4}. By repeating the previous argument, one 
obtains the whole chessboard aspect of the conductance schematically shown in 
lower part of Fig. \ref{fig4}.

Since Kondo-like behavior only occurs when the experimental temperature 
is in the range between $(\Delta E)^2/T_K$ and $T_K$, 
a clear prediction of our scheme is that the highly conducting regions of the 
chessboard would become narrower for decreasing temperature due to the 
lower limit condition.

\section{Summary}
We present a theoretical analysis of the chessboard aspect of the 
conductance through a QD in the Kondo regime. 
We perform a numerical calculation of the spectrum of N electrons in the isolated QD.
Two alternating types of GS are found: strongly correlated and weakly 
correlated. By projecting onto the subspace of the two lowest eigenstates and
coupling to the leads, we get an effective Kondo Hamiltonian with antiferromagnetic 
exchange couplings depending on the QD properties. 
The main result of our description is the appearance of strong oscillations of 
the Kondo temperature due to strong oscillations
of the spectral amplitude for strongly and weakly correlated states.
This explains the chessboard oscillations of the experimental conductance measured 
at a fix temperature.

We make two predictions that can be experimentally checked:

1- The application of an inplane magnetic field provokes the transition from 
the regime of strong correlation to that of weak correlation. 
As a consequence, 
one should detect a clearly measurable increase of $T_K$ during this process.

2- Since Kondo-like behavior only occurs when the experimental temperature 
is in the range $(\Delta E)^2/T_K<T<T_K$, the highly conducting regions of the 
chessboard would become narrower for decreasing temperature due to the 
lower limit condition.

\begin{acknowledgements}
We are grateful to J. Weis for useful discussions and providing us with 
experimental information.
This work was supported in part by MEC (Spain) under contract No. PB96-0085,
and CAM (Spain) under contract No. 07N/0064/2001. 
\end{acknowledgements}

{} 

\end{document}